\documentclass[journal, twocolumn, 10pt]{IEEEtran}

\usepackage[pdftex]{graphicx}
\usepackage[cmex10]{amsmath,empheq}
\usepackage{cite}
\usepackage{url}
\usepackage{color}
\usepackage{microtype}
\usepackage{amssymb}
\usepackage{relsize}
\usepackage{lipsum}
\usepackage{epstopdf}
\usepackage[font = footnotesize]{caption}
\usepackage{enumitem}
\usepackage{bm}
\usepackage{multirow}
\usepackage{array}
\usepackage{ifthen}
\usepackage{float}
\usepackage{subfig}
\usepackage{xcolor}% http://ctan.org/pkg/xcolor
\usepackage{algpseudocode}%http://ctan.org/pkg/algorithmicx
\usepackage{algorithm}% http://ctan.org/pkg/algorithm
\usepackage{epstopdf}

\algnewcommand{\algorithmicand}{\textbf{ and }}
\algnewcommand{\algorithmicor}{\textbf{ or }}
\algnewcommand{\OR}{\algorithmicor}
\algnewcommand{\AND}{\algorithmicand}
\algnewcommand{\var}{\texttt}

\begin{document}
\relscale{0.98}

\title{Frequency-Selective PAPR Reduction for OFDM}

\author{Selahattin G\"{o}kceli,
		Toni Levanen,
		Taneli Riihonen,
		Markku Renfors,
        and~Mikko Valkama %\vspace{-7pt}% <-this % stops a space
%\thanks{The research work leading to these results was funded in part by Nokia Bell Labs and in part by the Finnish Funding Agency for Innovation (Tekes, under the project ``Wireless for Verticals, WIVE'') (\emph{Corresponding Author: Mikko Valkama}).}%
%\thanks{Manuscript received Nov. 5th, 2018; revised on Feb. 2nd, 2019.}
\thanks{S. G\"{o}kceli, T. Levanen, T. Riihonen, M. Renfors, and M. Valkama are with 
Tampere University, Tampere, Finland (\emph{Corresponding Author:} Mikko Valkama, e-mail: mikko.valkama@tuni.fi).

© 2019 IEEE.  Personal use of this material is permitted.  Permission from IEEE must be obtained for all other uses, in any current or future media, including reprinting/republishing this material for advertising or promotional purposes, creating new collective works, for resale or redistribution to servers or lists, or reuse of any copyrighted component of this work in other works.

This work has been accepted for publication as a Correspondence in the IEEE Transactions on Vehicular Technology. Copyright may be transferred without notice, after which this version may no longer be accessible.}}

% The paper headers
%\markboth{IEEE Transactions on Vehicular Technology, Correspondence Paper, revised on Feb. 2nd, 2019}%
{}

% make the title area
\maketitle

% As a general rule, do not put math, special symbols or citations
% in the abstract or keywords.
\begin{abstract}
We study the peak-to-average power ratio (PAPR) problem in orthogonal frequency-division multiplexing (OFDM) systems. In conventional clipping and filtering based PAPR reduction techniques, clipping noise is allowed to spread over the whole active passband, thus degrading the transmit signal quality similarly at all active subcarriers. However, since modern radio networks support frequency-multiplexing of users and services with highly different quality-of-service expectations, clipping noise from PAPR reduction should be distributed unequally over the corresponding physical resource blocks (PRBs). To facilitate this, we present an efficient PAPR reduction technique, where clipping noise can be flexibly controlled and filtered inside the transmitter passband, allowing to control the transmitted signal quality per PRB. Numerical results are provided in 5G New Radio (NR) mobile network context, demonstrating the flexibility and efficiency of the proposed method.
\end{abstract}

%\vspace{-1pt}
% Note that keywords are not normally used for peerreview papers.
\begin{IEEEkeywords}
5G New Radio (NR), clipping, filtering, orthogonal frequency-division multiplexing (OFDM), peak-to-average-power ratio (PAPR), waveform, wireless communications.
\end{IEEEkeywords}

\section{Introduction}
\label{sec:intro}
\IEEEPARstart{5}{G} New Radio (NR) is expected to bring significant improvements over existing systems in data rates, reliability, latency and energy consumption \cite{NR.300}, \cite{DAHLMAN201857}, with physical layer radio access building on the cyclic prefix (CP) orthogonal frequency division multiplexing (OFDM). In order to preserve the benefits of OFDM, its peak-to-average power ratio (PAPR) behavior should be rendered compatible with available hardware characteristics and transmit signal quality requirements. As detailed in \cite{Jiang}, low PAPR is crucial to improve the transmitter power efficiency and to reduce the power amplifier (PA) related signal degradation and unwanted emissions. The main available methods addressing the PAPR problem are iterative clipping and filtering (ICF) \cite{clipping}, partial transmit sequence \cite{pts}, selected mapping \cite{SLM}, and tone reservation (TR) \cite{7024183}, with good overview being available in \cite{taxonomy}. 

In ICF that is one of the most efficient PAPR reduction methods, clipping and filtering are iteratively applied to obtain the desired PAPR behavior while suppressing the clipping-induced out-of-band emissions \cite{ICF1}. The filtering is commonly implemented by weighting the subcarriers (SCs) of the clipped signal in frequency domain. When low PAPR levels are targeted, conventional ICF is known to require a fairly large number of iterations. Therefore, improved schemes have been proposed \cite{ICF2}, \cite{ICF3} to decrease the number of iterations, however, at the cost of extra computational complexity at individual iteration. 

In modern cellular networks, primarily 3GPP Long Term Evolution (LTE)/LTE-Advanced and 5G NR, aggressive frequency-domain multiplexing of users with highly different quality-of-service expectations is utilized \cite{NR.300}, \cite{DAHLMAN201857}. Additionally, the adopted modulation and coding scheme (MCS) can be controlled at physical resource block (PRB) level, reflecting the instantaneous channel qualities across frequencies.  Therefore, the clipping noise power should be distributed unequally over the active PRBs, unlike in conventional solutions, where the clipping noise is spreading evenly over the whole transmitter passband. 

In the existing literature, the idea of controlling the clipping noise inside the transmitter passband has been pursued only to a limited extent.
In \cite{Fehri} and \cite{Traverso}, filter design methods are %for unequal distribution of clipping noise is 
discussed and pursued, primarily in the context of carrier aggregation transmitters, to allow for different passband clipping noise levels at different component carriers or bands. Additionally, the TR method basically reserves specific subcarriers for clipping-induced distortion to keep the actual data subcarriers free from clipping noise. In \cite{TR}, PAPR reduction is modeled as a peak cancellation task and an efficient peak cancellation signal generation method is proposed for TR. However, the TR approach is fundamentally different from ICF based methods in the sense that the reserved tones do not carry any data and represent thus substantial overhead in the overall system design.

In this correspondence, we propose an efficient and flexible PAPR reduction solution, referred to as iterative clipping and error filtering (ICEF), which allows for PRB-level control of the involved clipping noise. Different to existing solutions, the proposed method is based on explicitly separating the clipping noise in frequency domain, within the overall iterative procedure, and adopting a frequency domain mask to control the noise power at different passband PRBs as well as at the stopband regions. With proper frequency domain mask, even clipping noise -free passband PRBs can be obtained, which can be beneficial, e.g., for adopting the highest MCSs or improving the reliability in ultra-reliable low-latency communications (URLLC) applications \cite{3GPPTS22261}. This approach can be seen to have some similarity also to TR based methods, however, in the proposed concept no extra overhead is imposed and all passband PRBs can carry data. As verified with extensive numerical examples in 5G NR context, clipping noise free passband PRBs can be supported, while simultaneously being able to efficiently control the PAPR, without utilizing complicated optimization procedures or knowledge of the input signal's distribution. Finally, the proposed ICEF method has very similar computational complexity as the original ICF method, as detailed in the paper.

\section{System Model and Proposed Solution}
\label{sec:system_model}

%%% commenting out FIG.1 for now ...
\begin{figure}[tb]
	\centering
	\includegraphics[width=2.7in]{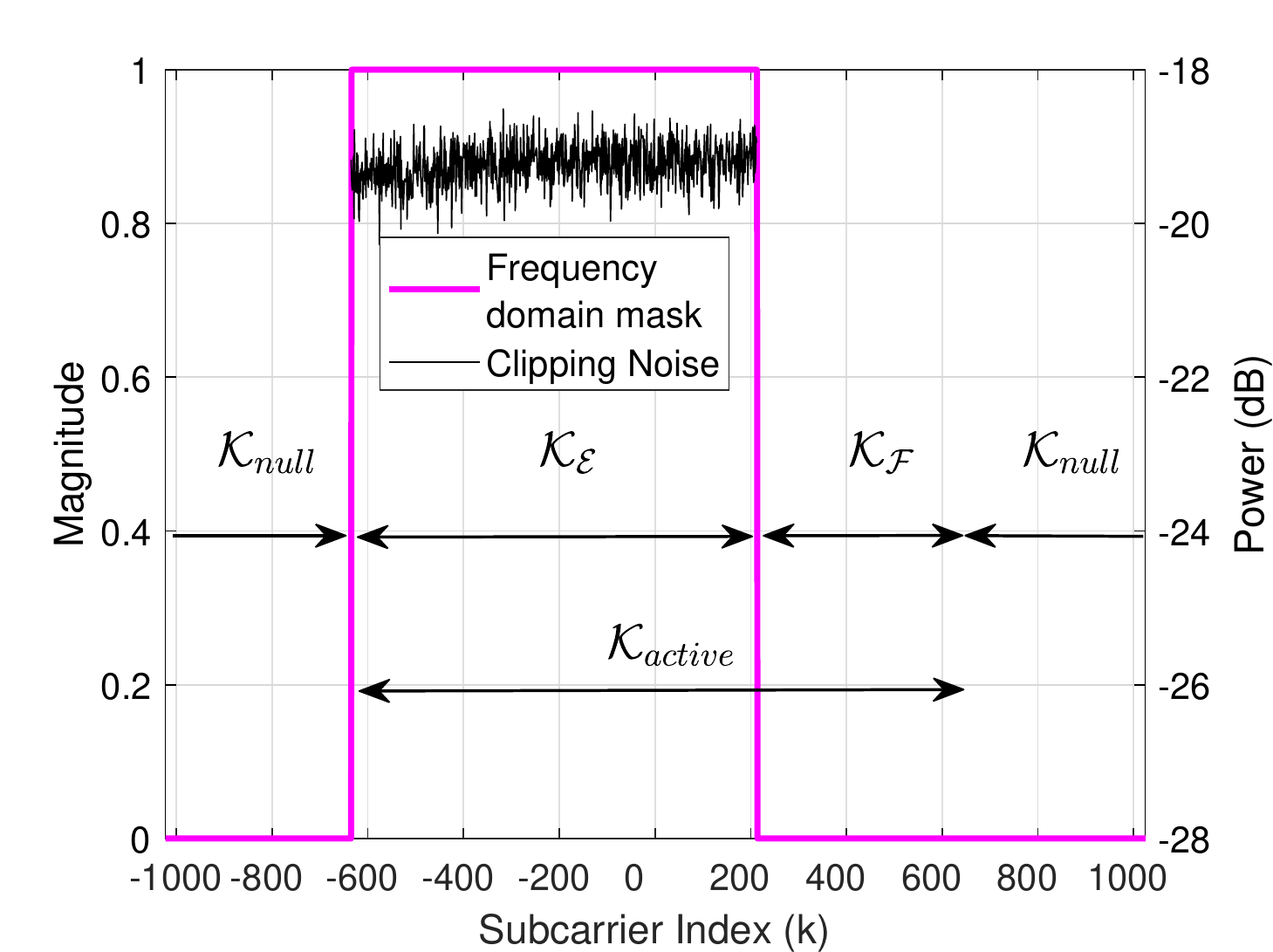}
	\caption{Basic illustration of a frequency domain mask allowing clipping noise only at active SCs $-636...212$. In general, the distorted, clean and non-active PRB/SC regions are denoted by $\mathcal{K_E}$, $\mathcal{K_F}$ and $\mathcal{K}_{null}$, respectively, while the total active passband is $\mathcal{K}_{active}=\mathcal{K_E} \cup \mathcal{K_F}$.} 
	\label{w1}
%	\vspace*{-0.4cm}
\end{figure} 

\subsection{Basics and Traditional ICF}

We denote the nominal transform size of the OFDM waveform processing by $N_{DFT}$. We assume that only $N_{act}<N_{DFT}$ of the involved subcarriers are active, which for simplicity are assumed to be located symmetrically around the DC bin when double-sided bin and subcarrier indexing is considered. In addition, $N=N_{DFT}N_{ov}$ represents the oversampled IDFT size, where $N_{ov}$ is the oversampling factor. Oversampled transforms are commonly needed to reliably model the peaks of the corresponding true analog waveform \cite{975762}.
Then, the original discrete-time OFDM symbol samples, reflecting the  $0^\textrm{th}$ iteration in iterative clipping and filtering methods, can be expressed as
\begin{align}
{x}^0[n]=\frac{1}{\sqrt[]{N_{act}}}\sum_{k=-N_{act}/2}^{N_{act}/2-1}X^0[k]e^{j2\pi kn/N}, 
{\rm }n = 0, \ldots, N-1,
\end{align}
where $X^0[k]$ represents the data symbol at active subcarrier $k \in \{-N_{act}/2, ..., N_{act}/2-1\}$ while all other IDFT bins are assumed to be stacked with zeros, and $n$ corresponds to the relative sample index inside the OFDM symbol.

In the continuation, we use $l$ to denote the iteration index in the PAPR reduction methods, and $x^{l}[n]$ the corresponding time-domain OFDM signal. The PAPR of $x^{l}[n]$ is defined as
\begin{align}
    \text{PAPR}(\mathbf{x}^l) =  10\log_{10}\frac{\max_{ n=0,1,\ldots,N-1}\big\{|x^l[n]|^2\big\}}{\frac{1}{N} \sum_{n=0}^{N-1}\{|x^l[n]|^2\} }, 
    \label{PAPRcalcu}
\end{align}
where $\text{max}\{\cdot\}$ represents maximum operator. The PAPR target is denoted by $\lambda_{\text{target}}$ while maximum iteration count is $L$. The soft limiter based clipping operation is defined as
\begin{align}
    \bar{x}^l[n] = \begin{cases}
   Ae^{\angle x^{l-1}[n]}, & \text{if} \; |x^{l-1}[n]|>A\\
   x^{l-1}[n], & \text{otherwise}
\end{cases}
\label{clipping}
\end{align}
where $\angle x$ and $|x|$ denote the phase angle and modulus of a complex number $x$, respectively, while $\bar{x}^l[n]$ and $A$ represent the clipped version of $ x^{l-1}[n]$ and the amplitude threshold value which is computed in accordance with the target PAPR level, respectively. The clipping function in (3) is deliberately defined, for presentation convenience, to increase the iteration index by one in the context of iterative methods. 

In traditional ICF \cite{clipping}, each iteration starts with above clipping operation which is followed by a DFT, of size $N$, and frequency domain filtering in order to suppress the clipping noise outside the passband region. To express this more formally, we define the ICF frequency domain filter mask as
\begin{align}
H_{\text{ICF}}[k] = \begin{cases}
1, & \text{if} \; k\in\mathcal{K}_{active} \\
0, & \text{if} \; k\in\mathcal{K}_{null}
\end{cases}
\label{equICF}
\end{align}
where $\mathcal{K}_{active}$ and $\mathcal{K}_{null}$ represent the sets that contain the active SCs and non-active SCs, respectively, and the cardinality of the union of the sets equals to $|\mathcal{K}_{active} \cup \mathcal{K}_{null}|=N$. The frequency-domain filtering operation at iteration $l$ reads then
\begin{align}
X^{l}[k] = H_{\text{ICF}}[k]\bar{X}^{l}[k],
\label{trad2}
\end{align}
and the corresponding time-domain signal $ x^{l}[n]$ is obtained through IDFT of size $N$. As is evident from (4)-(\ref{trad2}), the passband clipping noise is not processed and thus all active PRBs are subject to similar distortion. Furthermore, since the frequency domain filter is applied directly on the total clipped signal, the degrees of freedom to tailor the filter mask at passband, compared to all ones shown in (4), are very limited to avoid creating linear distortion in the transmitted signal. This fact generally limits the flexibility and the passband noise suppression gains that can be obtained with the traditional ICF approach.

\subsection{Proposed Method} %{\color{red} ..will cont from here tomorrow}
The proposed ICEF method builds on the idea of calculating or separating the prevailing clipping noise in each iteration. Through this approach, we can more freely and flexibly choose the frequency-domain weight window or mask which is thereon used to filter only the clipping noise. This, in turn, allows us, e.g., to concentrate the clipping noise into specific PRBs only while leaving others fully clipping noise free. In this approach, the filtered clipping noise, when transformed to time-domain, can also be interpreted as an approximate peak cancellation signal.

We define the frequency-domain mask of the proposed method as
\begin{align}
H_{\text{ICEF}}[k] = \begin{cases}
1, & \text{if} \; k\in\mathcal{K_E} \\
0, & \text{if} \; k\in\mathcal{K_F} \cup \mathcal{K}_{null},
\end{cases}
\label{equH}
\end{align}
where $\mathcal{K_E}$ and $\mathcal{K_F}$ represent the mutually exclusive sets of passband SCs where clipping noise is allowed and not allowed, respectively, while $\mathcal{K}_{null}$ is as defined earlier. It thus follows that $\mathcal{K}_{active}=\mathcal{K_E} \cup \mathcal{K_F}$, while $\mathcal{K_E} \cap \mathcal{K_F} = \O$, with an example graphical illustration shown in Fig. \ref{w1}. The clipping noise at iteration $l$ can be obtained in frequency-domain as 
\begin{align}
C^{l}[k] = \bar{X}^{l}[k] - X^0[k].
\label{equc}
\end{align} 
This is then filtered with the mask defined in (6) and added back to the original non-clipped signal, expressed in frequency-domain as 
\begin{align}
X^{l}[k] = X^0[k]+H_{\text{ICEF}}[k]C^{l}[k].   
\label{last}
\end{align}
Finally, the actual time-domain signal ${x}^{l}[k]$ is obtained through IDFT of size $N$. The overall processing flow is summarized in Algorithm \ref{alg:ICEF} depicting the iterative processing executed for every OFDM symbol, while using vector notation for presentation convenience and compactness. The subsequent processing stages, such as CP addition, are implemented as in any OFDM transmitter.

As the most important feature, the filtered clipping noise or peak cancellation signal is processed separately, as an additive element to the initial OFDM signal.
%from the actual linear data signal. 
The conventional ICF can be interpreted as a special case where $\mathcal{K_F} = \O$ and thus the clipping noise is distributed over all active SCs. In ICEF, however, $\mathcal{K_F}$ is not empty and as shown in (6), can be defined to include clean PRBs which do not contain any clipping noise. It should also be noted that if the overall passband width (size of $\mathcal{K}_{active}$) is constant, then the ICF mask is basically fixed as shown in (4). However, in case of ICEF, different frequency-domain masks can be flexibly generated based on, e.g., the scheduling decisions and link adaptation information, which thus brings large additional flexibility in the system optimization. Importantly, we also note that the proposed ICEF principle is not limited to the basic mask definition in (6), with only binary levels inside the passband. Thus, yet more versatile masks with, e.g, finite suppression of the clipping noise in selected PRBs can be flexibly defined. More elaborate treatment of such scenarios is, however, outside the scope of this paper, forming an important topic for future work.

\subsection{Computational Complexity}
The computational complexities of the original ICF and the proposed ICEF methods are generally very similar. More specifically, it follows from (6)-(8) that the proposed ICEF method does not need any additional arithmetic operations, compared to the original ICF, if purely binary weights as shown in (6) are used. This is because the generation of the frequency-domain samples $X^l[k]$ in (8) can in practice be directly realized as 
\begin{align}
X^{l}[k] = \begin{cases}
\bar{X}^{l}[k], & \text{if} \; k\in\mathcal{K_E} \\
X^{0}[k], & \text{if} \; k\in\mathcal{K_F}\\
0, & \text{if} \; k\in\mathcal{K}_{null}.
\end{cases}
\label{equBinMaskMapping}
\end{align}
We note that (\ref{equBinMaskMapping}) applies only to the case with binary weights defined in (6), while if more arbitrary weights are used then some extra multiplications and additions are needed.

To shortly address the generalized case with arbitrary real-valued weights, we quantify the overall processing complexity of ICEF per iteration, as well as that of ordinary ICF for comparison purposes, including also the FFT and IFFT operations that are inherent and common to both methods. To this end, acccording to \cite{J:1987_Sorensen_FFT_complexity}, when the split-radix FFT algorithm is deployed, the computation of an $N=2^M$ point complex FFT requires $MN-3N+4$ real multiplications and $3MN-3N+4$ real additions. Both the ICF and ICEF require computation of two FFTs which doubles the given number of operations. Then in case of the proposed ICEF, the maximum complexity increase corresponds to the case of using non-trivial real weights for every frequency bin. In this case, the proposed method introduces $2N$ real additions to realize the separation of clipping noise (7), and another $2N$ real additions to combine the weighted clipping noise with the original signal (8). Furthermore, the weighting of the clipping noise in (8) requires $2N$ real multiplications, assuming real-valued weights. Therefore, the maximum computational complexity increase per iteration for the proposed ICEF method, in its most general form, corresponds to $2N$ real multiplications and $4N$ real additions. If the values shown in Table I are considered as a concrete example case, these correspond to approximately 9.1\% increase in the number of real multiplications and approximately 5.1\% increase in the number of real additions, compared to ordinary ICF.

\algnewcommand{\algorithmicgoto}{\textbf{Go to step}}%
\algnewcommand{\Goto}[1]{\algorithmicgoto~\ref{#1}}%
\begin{algorithm} [!t]
  \caption{Proposed ICEF algorithm} \label{icefalgo}
  \begin{algorithmic}[1]
 %   \Procedure{ICEF}{$x^0,\lambda_{Target}, L$}
    \State Set $\mathcal{K_E}$, $\mathcal{K_F}$, $\mathcal{K}_{null}$ and create $H_{\text{ICEF}}[k]$ according to (\ref{equH})
    \State Set $l=0$
    \State \label{marker1}Compute $\textrm{PAPR}(\mathbf{x}^{l})$ according to (\ref{PAPRcalcu}) 
    \If{$(\textrm{PAPR}(\mathbf{x}^{l}) > \lambda_{\text{target}})$ \AND  $(l < L)$} 
        \State Set $l = l+1$
        \State Compute $\bar{\mathbf{x}}^l$ according to (\ref{clipping})
        \State Compute $\bar{\mathbf{X}}^l=\textrm{DFT}\{\bar{\mathbf{x}}^l\}$
        \State Compute $\mathbf{C}^{l}$ according to (\ref{equc})
        \State Compute $\mathbf{X}^l$ according to (\ref{last})
        \State Compute $\mathbf{x}^l=\textrm{IDFT}\{\mathbf{X}^l\}$
        \State \Goto{marker1}
    \Else 
        \State {\Return $\mathbf{x}^l$}
    \EndIf
 \end{algorithmic}
\label{alg:ICEF}
\end{algorithm}

\begin{figure*} [tb]
	\centering
	\subfloat{\label{mseall}\includegraphics[width=0.31\linewidth]{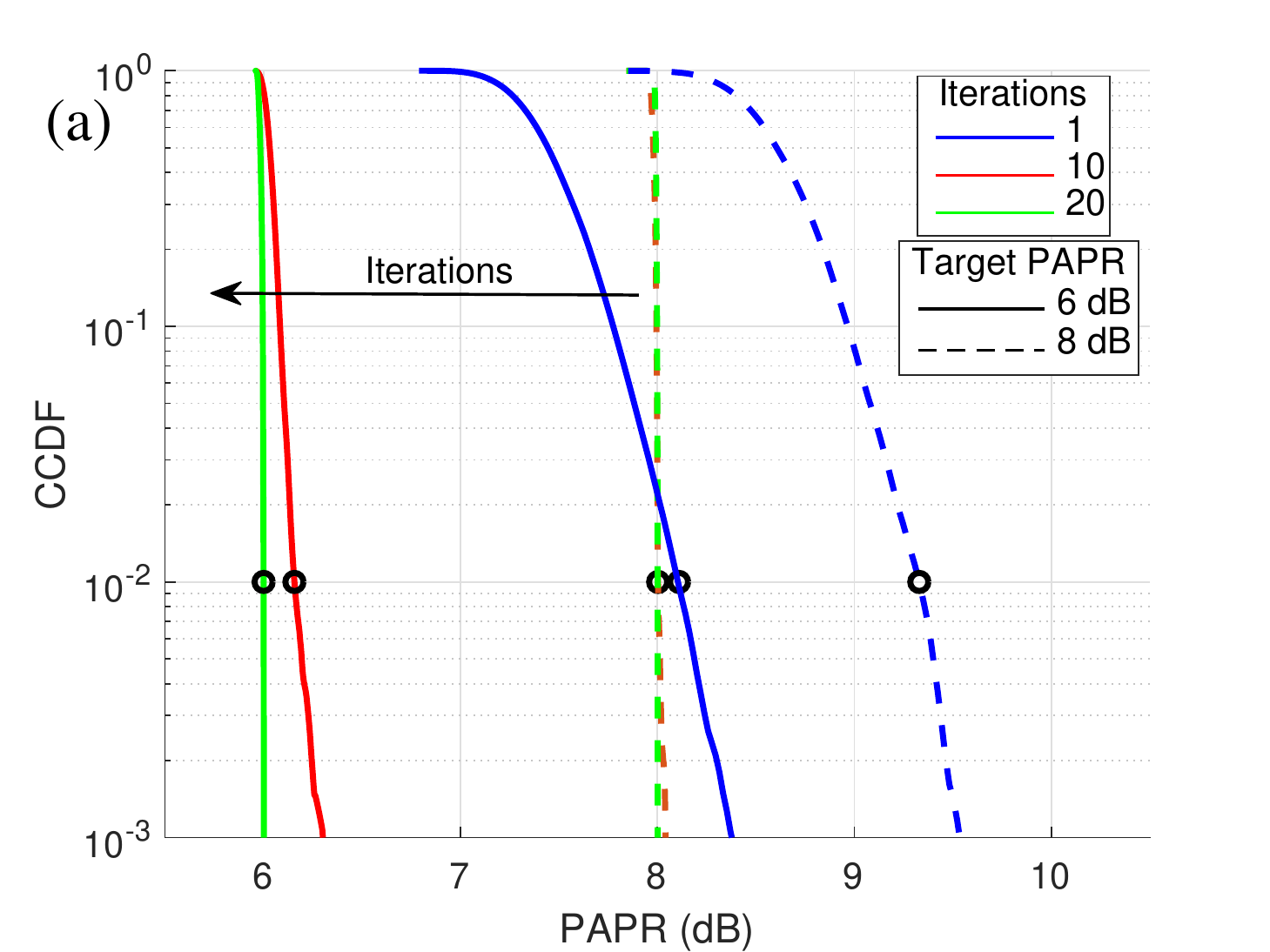}}
	\subfloat{\label{msepartial}\includegraphics[width=0.31\linewidth]{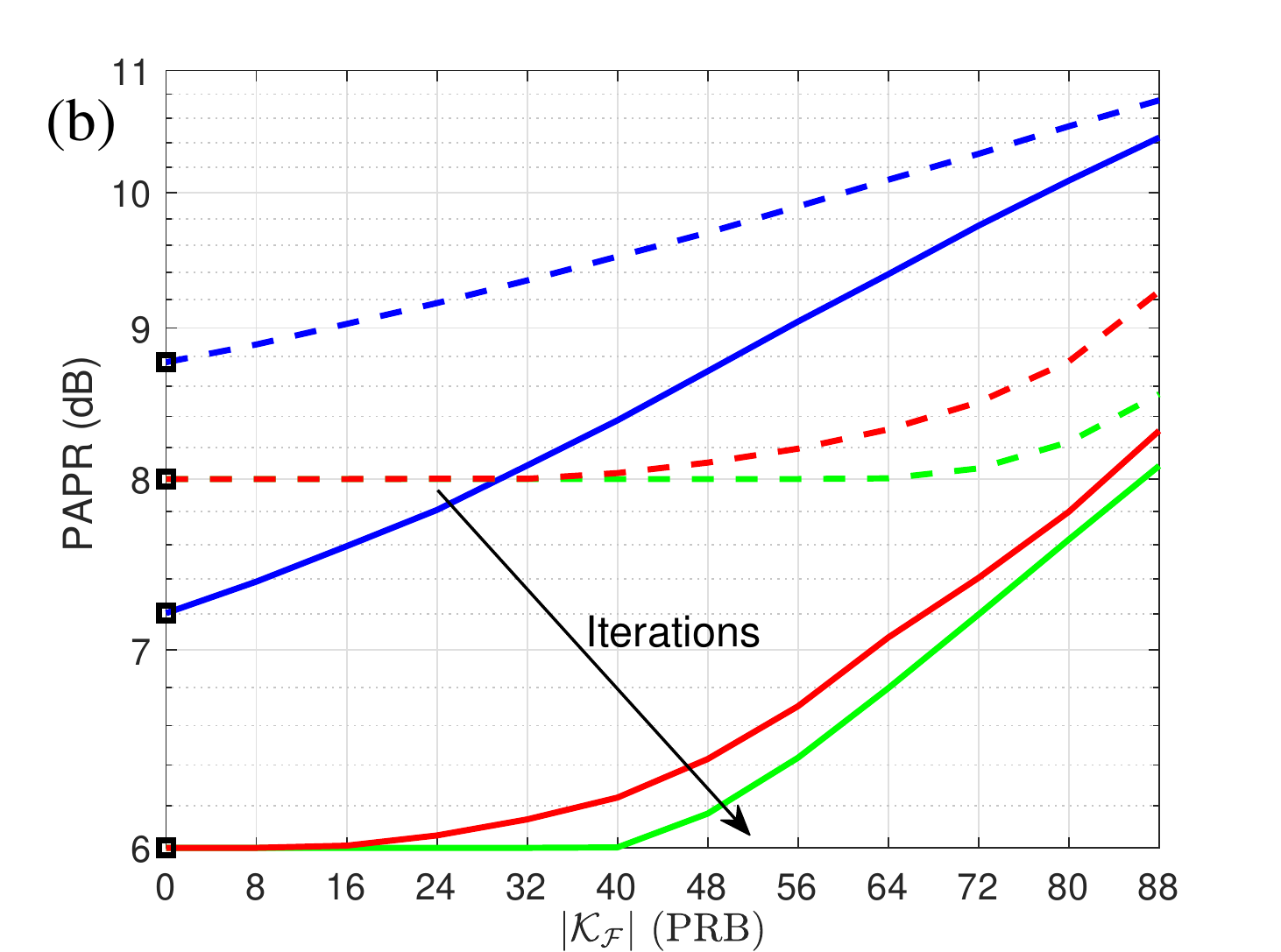}}
	\subfloat{\label{achievedPAPR}\includegraphics[width=0.31\linewidth]{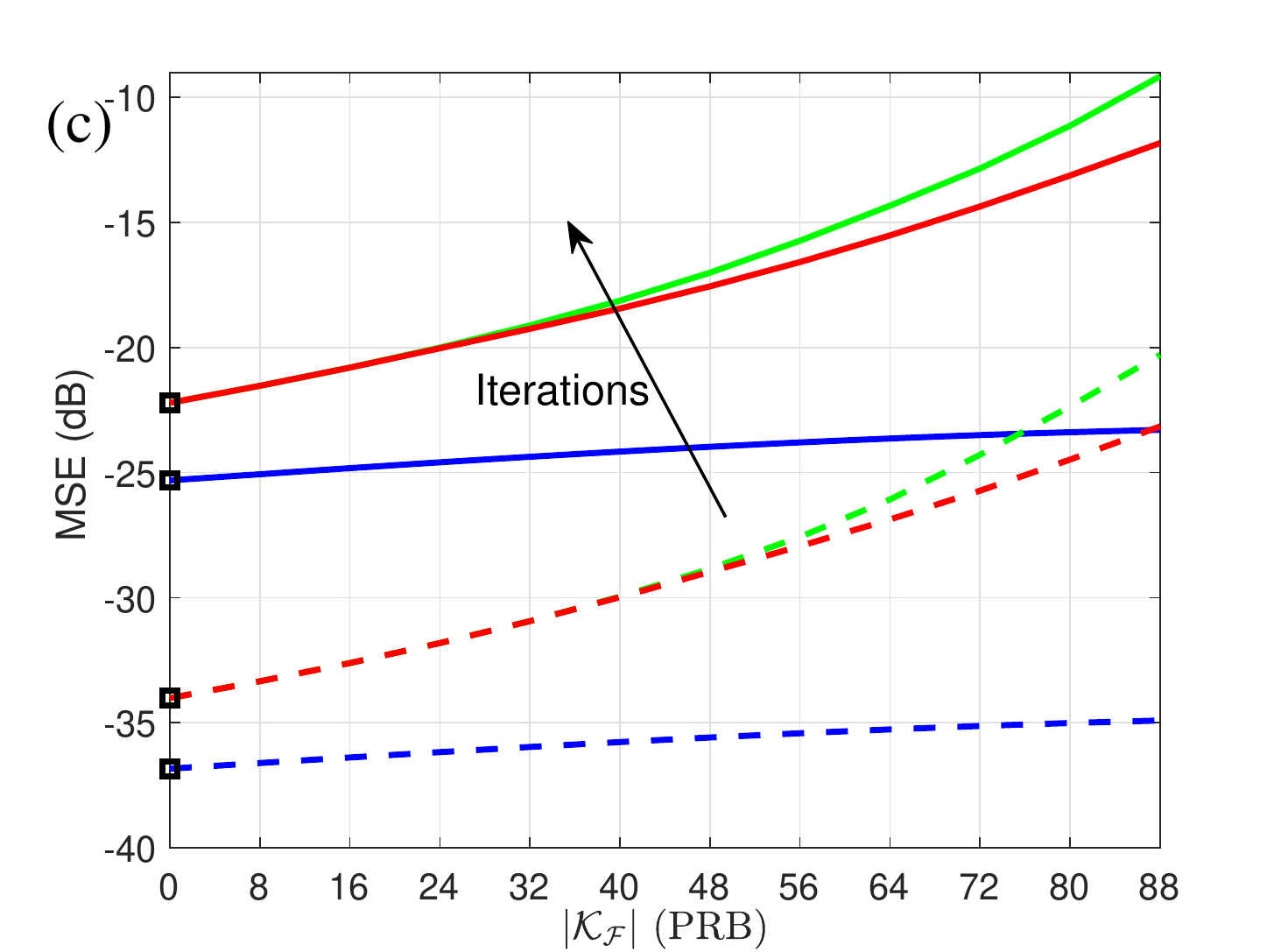}}
	\caption{PAPR and MSE results for different iteration counts and PAPR clipping target levels. In (a) PAPR distributions with $|\mathcal{K_F}|=34$ PRBs are shown. CCDF level of 1$\%$ is highlighted with black circles. For different sizes of the set $\mathcal{K_F}$, the achieved PAPR at 1$\%$ probability level and the MSE results for clipping noise distorted SCs are shown in (b) and (c), respectively. The traditional ICF algorithm performance results are also shown with black squares.
	%, for reference, for which $|\mathcal{K_F}|=0$.  
	\label{mse} 
	}
	%\vspace*{-0.4cm}
\end{figure*}

\section{Numerical Results and Analysis}
\label{sec:results}

\begin{table}[!t]
  \setlength{\tabcolsep}{3pt}
    \renewcommand{\arraystretch}{1.3}
    \footnotesize
    \centering
    \captionof{table}{Simulation and evaluation parameters.}
    %\caption{Test caption}
    \begin{tabular}{|l||l|}
    \hline
    \multicolumn{1}{|c||}{\textbf{Parameter}} & \multicolumn{1}{c|}{\textbf{Value}}\\
		\hline
        PRB size & 12 SCs \\ \hline
		Number of active SCs ($N_{act}$) / SC spacing & 1272 SCs / 15 kHz \\ \hline
		Nominal transform size ($N_{DFT}$) & 2048 \\ \hline
		Oversampling factor ($N_{ov}$) & 8  \\ \hline
		Modulation order & QPSK, 16-/64-/256-QAM \\ \hline
		Maximum number of iterations & 1, 10, 20 \\ \hline
		%test & 2 \\ \hline
    \end{tabular}
    \label{parameters}
\end{table}

In this section, numerical performance results are presented and analyzed. The evaluation scenarios are tailored to follow the 5G NR radio interface numerology defined in \cite{NR}, with the utilized main parameters shown in Table \ref{parameters}. Specifically, the number of active SCs is $N_{act}=1272$, or 106 PRBs with 12 SCs each, as defined in \cite{NR} for 20 MHz NR channel bandwidth. Complementary cumulative distribution function (CCDF) is used to quantify the PAPR reduction performance. It shows the probability that the envelope of the processed OFDM signal is above the level corresponding to a given PAPR value \cite{taxonomy}. Commonly, the achieved PAPR performance is numerically quantified at CCDF level of 1\% \cite{ICF2}, \cite{ICF3} and is the approach taken also here.

We first experiment the PAPR reduction performance with the size of the clean PRB set, denoted here as $|\mathcal{K_F}|$, ranging from 0 PRBs to 88 PRBs. The smallest size of clean PRB set, i.e., $|\mathcal{K_F}|=0$, corresponds to all PRBs being subject to clipping noise, and thus in this special case ICEF essentially merges with classical ICF. At the other extreme, only 18 PRBs contain clipping noise leaving thus 88 PRBs completely clipping noise free. 

For simple parameterization of this first experiment, high-pass like clipping noise mask is utilized and the number of clipping noise free PRBs is increased symmetrically starting from the PRBs at channel center towards the edge PRBs. In other words, at both edges there are $|\mathcal{K_E}|/2$ PRBs distorted by clipping noise and in the middle there are $|\mathcal{K_F}|$ clipping error free PRBs.

The obtained results are illustrated in Fig. \ref{mse}.  In Fig. \ref{mse}(a), the CCDFs are shown for an example case of $|\mathcal{K_F}|=34$ with two different target PAPR values and three different maximum iteration counts. Similar to ordinary ICF, it is observed that one iteration does not bring a good PAPR performance. However, with larger iteration numbers, very good PAPR behavior can clearly be obtained, despite the large share of the clipping noise free PRBs. In Fig. \ref{mse}(b), the achievable PAPR performance is shown, interpreted at CCDF probability level of 1\%, for different values of $|\mathcal{K_F}|$. With the maximum iteration count of 20, the given PAPR target of 6 dB or 8 dB can be strictly achieved while keeping up to 40 PRBs or 64 PRBs fully free from clipping noise, respectively. If we assume an additional tolerance of, e.g., 0.2 dB, the maximum numbers of clipping noise free PRBs increase further, to 48 PRBs and 78 PRBs. In Fig. \ref{mse}(c), the corresponding mean squared error (MSE) values are evaluated and shown for the distorted SCs while the SCs without clipping noise have no distortion. As expected, the MSE values at distorted PRBs increase when the amount of clean PRBs is increased. Additionally, the MSE values also increase when the target PAPR level is decreased from 8 dB to 6 dB. 

Interestingly, especially for the target PAPR level of 6 dB, acceptable MSE levels can be obtained even up to $|\mathcal{K_F}|=68$ clean PRBs. Specifically, as defined for 5G NR in \cite{NR}, $-15$~dB and $-18$~dB of MSE are required for QPSK and 16-QAM data modulations, respectively. Thus, assuming QPSK modulation and maximum iteration count of 10, any value of $|\mathcal{K_F}|$ between 0 to 68 PRBs meets this requirement for a target PAPR level of 6~dB. With 16-QAM, the corresponding range is 0 to 40~PRBs. %Considering also the effect of other transmitter impairments and allowing them to contribute one half of the error budget, the MSE performance results should be met with a 3~dB margin. Taking this into account, and interpreting the MSE results together with the previous PAPR results with the 0.2~dB tolerance and maximum iteration count of 20, we can finally conclude that up to $|\mathcal{K_F}|=40$ and $|\mathcal{K_F}|=16$ clean PRBs can be accommodated with QPSK and 16-QAM modulations, respectively, when the PAPR target is 6~dB. With 8~dB PAPR target, in turn, a maximum of $|\mathcal{K_F}|=78$ clean PRBs can be supported, for both QPSK and 16-QAM modulations. Overall, it can thus be observed and concluded that the proposed method is able to achieve very good PAPR reduction capability while simultaneously concentrating the clipping noise only at selected PRBs.

\begin{figure}[tb]
	\centering
	\includegraphics[width=2.7in]{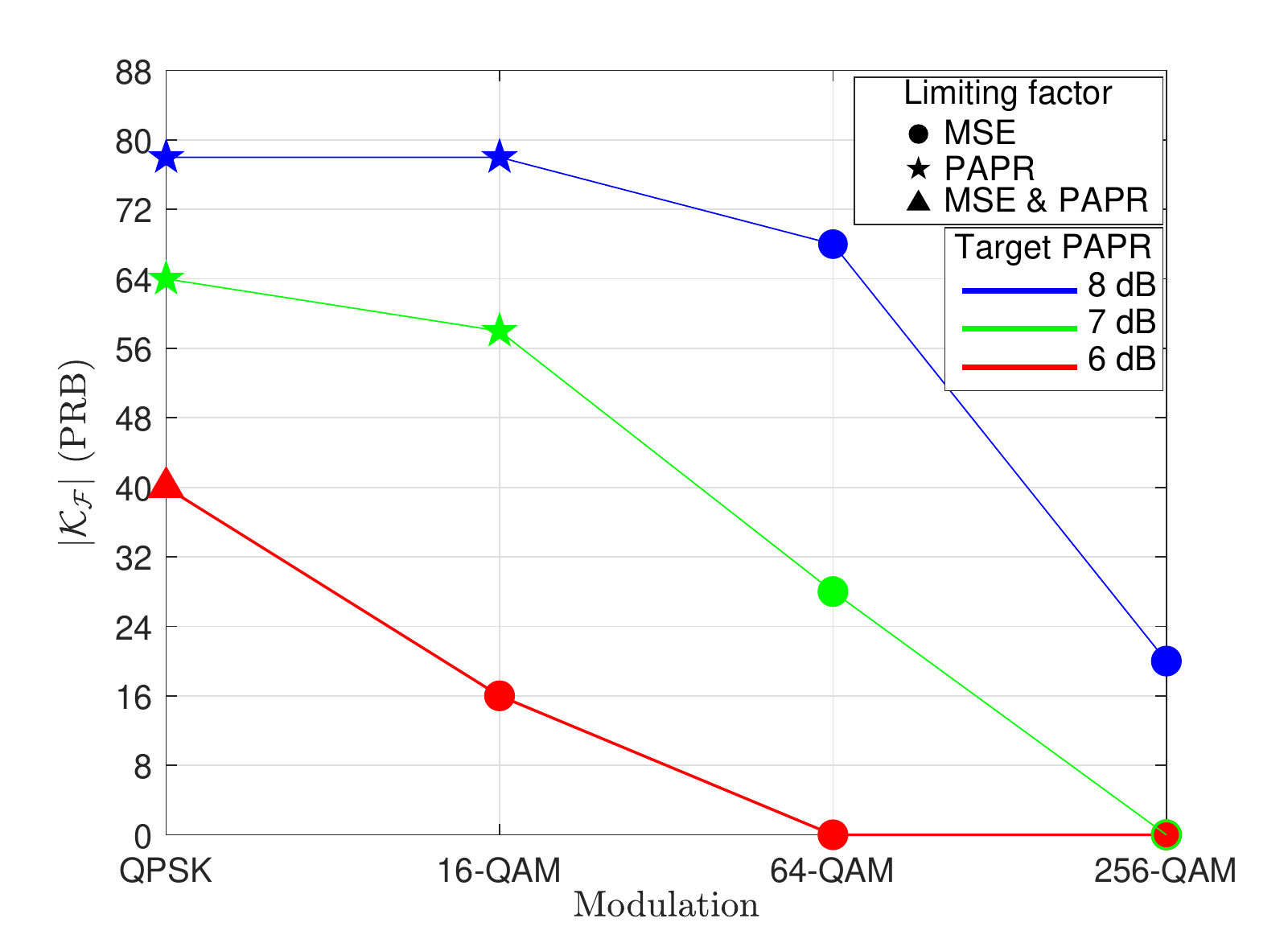}
	\caption{Maximum numbers of supported clipping-noise free PRBs for different PAPR targets and modulation specific MSE requirements. The $\mathcal{K_F}$ mapping follows the mapping described for Fig. \ref{mse}. On the x-axis, the modulation type assumed for subcarriers in $\mathcal{K_E}$ is given. 
	} 
	\label{kfMCS}
\end{figure}

Considering also the effect of other transmitter impairments and allowing them to contribute one half of the error budget, the MSE performance results should be met with a 3~dB margin. Taking this into account, and interpreting the MSE results together with the previous PAPR results with the 0.2~dB tolerance and maximum iteration count of 20, we evaluate and illustrate how the feasible size of the clipping noise free PRB set ($|\mathcal{K_F}|$) behaves with different modulation orders and target PAPRs. The results are shown in Fig. \ref{kfMCS}, indicating also what is the limiting factor (MSE, PAPR, or both) in different cases. When relatively high PAPR target levels such as 7~dB or 8~dB are assumed, the MSE requirements of QPSK and 16-QAM can be easily met. The number of error free PRBs is limited only by the PAPR target, and for both modulations $|\mathcal{K_F}|=78$ clipping error free PRBs can be provided with 8~dB PAPR target, and with 7~dB PAPR target QPSK and 16-QAM support $|\mathcal{K_F}|=64$ and $|\mathcal{K_F}|=58$ clipping error free PRBs, respectively. Then, when a lower PAPR level of 6 dB is targeted, maximum of $|\mathcal{K_F}|=40$ and $|\mathcal{K_F}|=16$ clipping error free PRBs can be supported with QPSK and 16-QAM, respectively. With 6~dB PAPR target the clipping distortion starts to increase significantly and the MSE becomes the main limiting factor with 16-QAM modulation. 

Additionally, Fig. \ref{kfMCS} shows also results assuming 64-QAM or 256-QAM modulation to be used in the PRBs where we allow clipping noise. These correspond to scenarios where 64-QAM or 256-QAM would be the lower modulation order while in the clipping free PRBs even higher modulation order would be applied, reflecting the general tendency of pursuing ever-increasing peak data rates and spectral efficiencies in the future wireless networks. The MSE levels allowed for 64-QAM and 256-QAM correspond to only -22~dB and -29~dB, respectively \cite{NR}. With 64-QAM modulation at the clipping error distorted subcarriers, $|\mathcal{K_F}|=68$ or $|\mathcal{K_F}|=28$ clipping error free PRBs can still be achieved when 7 dB or 8~~dB PAPR level are targeted, respectively. However, with 6~dB PAPR target no clipping error free PRBs can be supported. Moreover, even 256-QAM modulation can be supported at the clipping error distorted PRBs while providing $|\mathcal{K_F}|=20$ clipping error free PRBs for even higher order modulation when PAPR target level of 8 dB is assumed. Due to the strict MSE requirement of 256-QAM, no clipping error free PRBs can be supported with PAPR target levels of 6~dB or 7~dB. Overall, the proposed method is able to achieve very good PAPR reduction capability with different MSE requirements while simultaneously concentrating the clipping noise only at selected PRBs.

\begin{figure*} [htb]
	\centering
	\subfloat{%
		\includegraphics[width=0.33\linewidth]{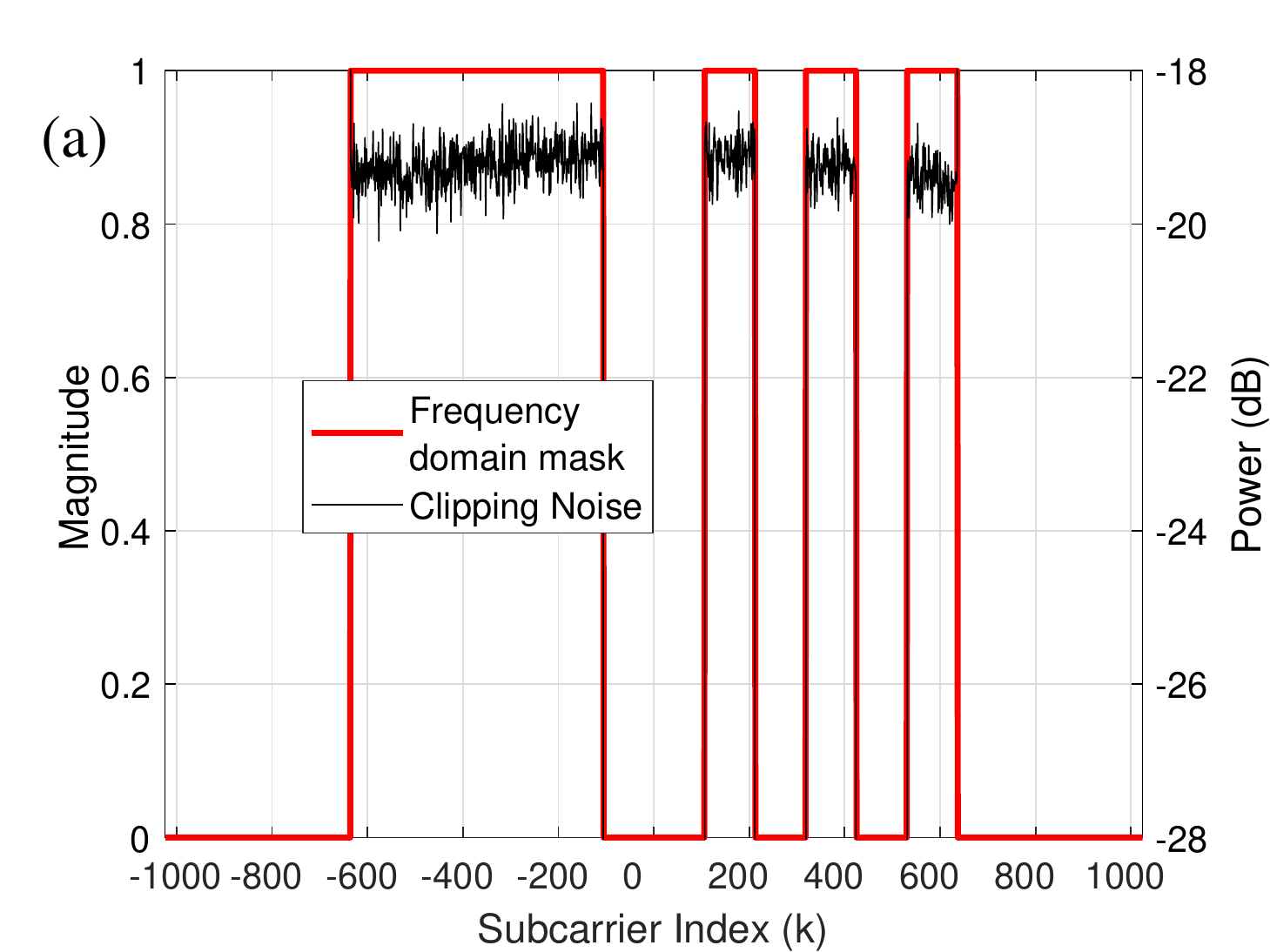}\label{w2}}
	\hfill
	\subfloat{%
		\includegraphics[width=0.33\linewidth]{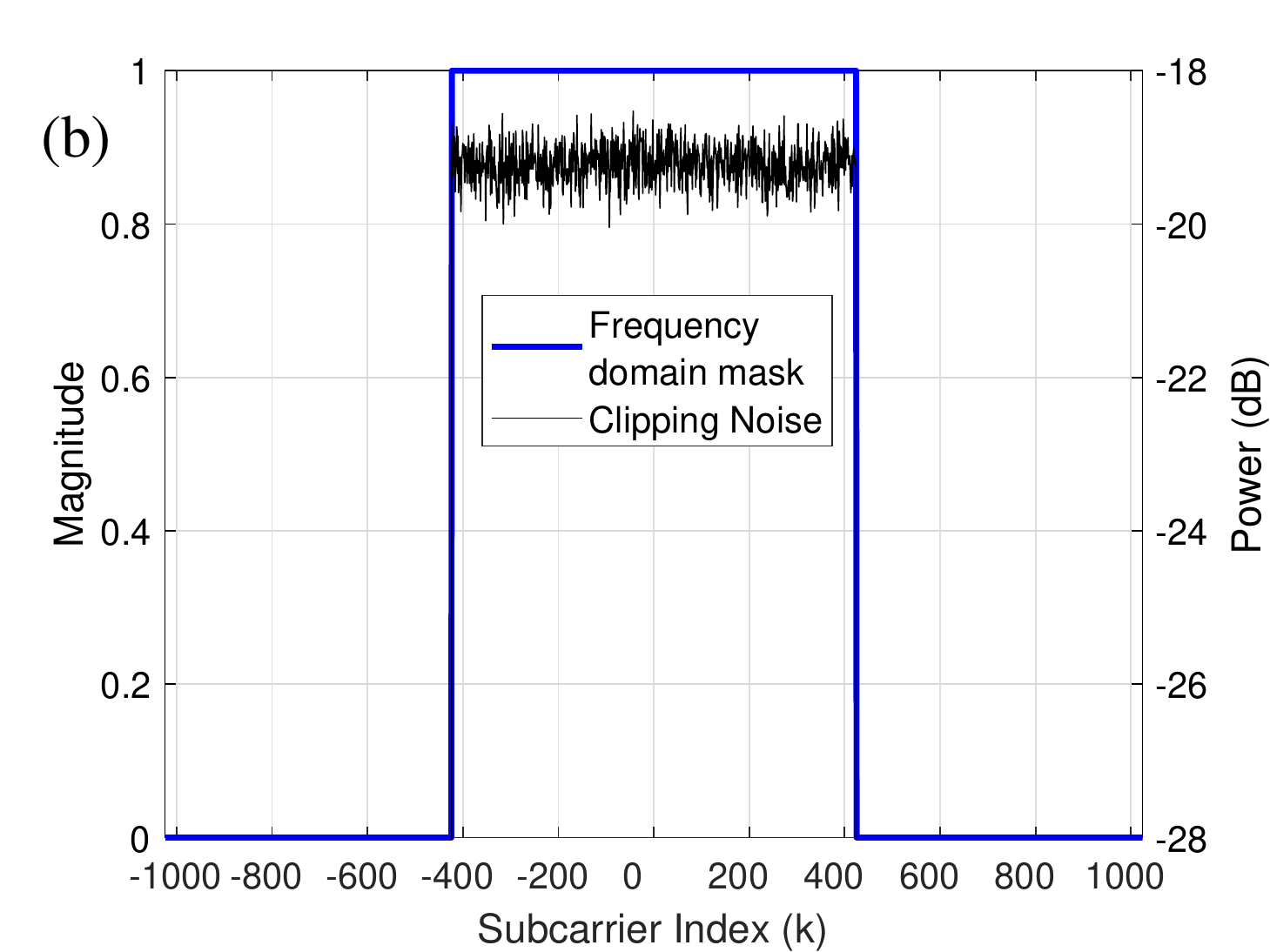}}
	\label{w3}\hfill
	\subfloat{%
		\includegraphics[width=0.33\linewidth]{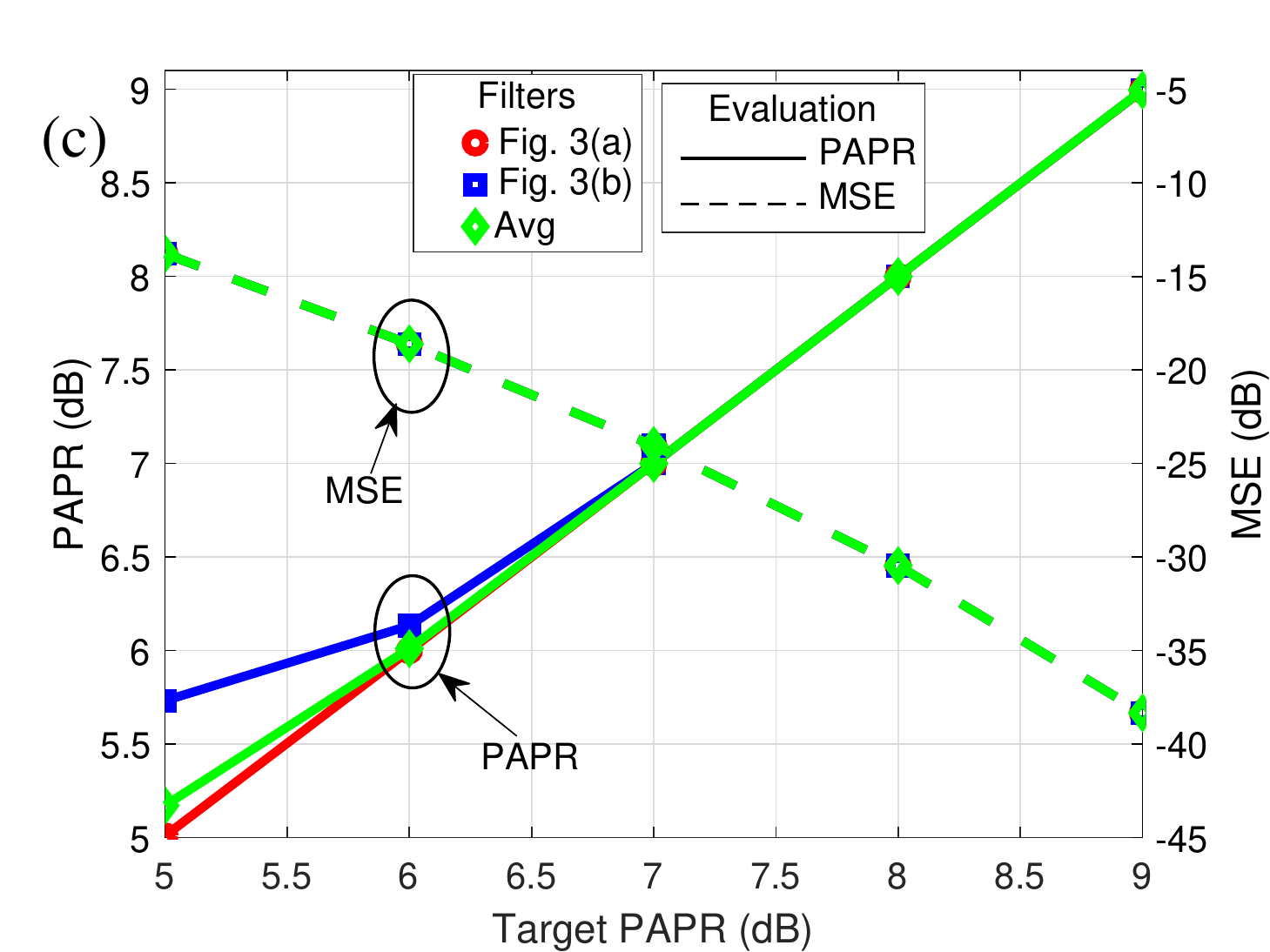}}
	\label{positionPAPR}
	\caption{Two example clipping noise masks are shown in (a) and (b), respectively. In (c), the corresponding PAPR performances are shown, together with the average performance over all masks, assuming 20 iterations. Additionally, also the MSE performance at distorted SCs is shown.} \label{position} 
\end{figure*}

As the other main evaluation case, the possible dependencies between the exact SC indexes in $\mathcal{K_F}$, the achievable PAPR reduction performance and the MSE behavior are next investigated. 

In these evaluations, the available 1272 active SCs are divided into 12 contiguous sub-bands, each containing 106 SCs. Then, out of these 12 sub-bands, eight are allowed to contain clipping noise while four are always noise-free, corresponding to $\mathcal{K_E}$ and $\mathcal{K_F}$, respectively. Then all possible combinations of the eight sub-bands
are simulated to understand the possible impact of the different ICEF mask types on the achievable PAPR performance.

The obtained results are shown in Fig. \ref{position}. First, the two specific noise mask realizations that result to the lowest and the highest PAPR values at $1\%$ CCDF level after 20 iterations are shown in Fig. \ref{position}(a) and \ref{position}(b), respectively, depicting also example snapshots of the clipping noise spectrum. The corresponding achieved PAPR levels with these two noise masks are shown in Fig. \ref{position}(c) together with the MSE values of the clipping noise distorted SCs, as functions of the target PAPR level. In addition, in Fig. \ref{position}(c), the average performance over all possible ICEF mask realizations is illustrated.
Importantly, it can be observed that the selection of the SCs or PRBs included in the set $\mathcal{K_F}$ does not cause any major difference in the PAPR performance, which brings good PRB allocation flexibility. It is also expected that different masks should provide a similar MSE performance. In line with this, as can be seen in Fig. \ref{position}(c), the MSE performance is the same for all considered cases. Thus, the user scheduling and their PRB allocation can follow the instantaneous channel quality indicators or measurements, without imposing any constraints on the ICEF mask. The proposed frequency-selective PAPR reduction method can then be used to improve the efficiency of the MCS selection or the link reliability, independent of the exact scheduling and PRB allocation decisions.

\section{Conclusions}
\label{sec:conclusions}

In this paper, we have presented a novel frequency-selective PAPR reduction solution for OFDM systems to facilitate different transmit signal quality levels at different PRBs. The proposed iterative clipping and error filtering (ICEF) method builds on the idea of explicitly separating the prevailing clipping noise, at every iteration, and adopting a frequency domain mask to control the distortion at different PRBs. The proposed method was evaluated in 5G NR system context through extensive simulations. Based on the obtained results, the proposed method is able to efficiently control the PAPR while simultaneously facilitating a large share of distortion-free PRBs. Additionally, the results show that the method is robust against the exact allocations of the clean PRBs, offering thus a flexible solution to, e.g., improve the radio link reliability for URLLC devices or further increase the MCS for high throughput users. Our future work will focus on more elaborate design and optimization of the frequency-domain ICEF mask for different transmitter scenarios and applications.

\bibliographystyle{IEEEtran}
\bibliography{main}

% Generated by IEEEtran.bst, version: 1.12 (2007/01/11)
\begin{thebibliography}{10}
\providecommand{\url}[1]{#1}
\csname url@samestyle\endcsname
\providecommand{\newblock}{\relax}
\providecommand{\bibinfo}[2]{#2}
\providecommand{\BIBentrySTDinterwordspacing}{\spaceskip=0pt\relax}
\providecommand{\BIBentryALTinterwordstretchfactor}{4}
\providecommand{\BIBentryALTinterwordspacing}{\spaceskip=\fontdimen2\font plus
\BIBentryALTinterwordstretchfactor\fontdimen3\font minus
  \fontdimen4\font\relax}
\providecommand{\BIBforeignlanguage}[2]{{%
\expandafter\ifx\csname l@#1\endcsname\relax
\typeout{** WARNING: IEEEtran.bst: No hyphenation pattern has been}%
\typeout{** loaded for the language `#1'. Using the pattern for}%
\typeout{** the default language instead.}%
\else
\language=\csname l@#1\endcsname
\fi
#2}}
\providecommand{\BIBdecl}{\relax}
\BIBdecl

\bibitem{NR.300}
{3GPP TS 38.300 V15.4.0}, ``{New Radio (NR); Overall description; Stage-2},''
  \emph{{\normalfont Tech. Spec. Group Radio Access Network, Rel. 15”}}, Dec.
  2018.

\bibitem{DAHLMAN201857}
E.~Dahlman, S.~Parkvall, and J.~Sk\"{o}ld, \emph{{5G {NR}: the Next Generation
  Wireless Access Technology}}.\hskip 1em plus 0.5em minus 0.4em\relax Academic
  Press, 2018.

\bibitem{Jiang}
T.~Jiang and Y.~Wu, ``An overview: Peak-to-average power ratio reduction
  techniques for {OFDM} signals,'' \emph{IEEE Trans. Broadcast.}, vol.~54,
  no.~2, pp. 257--268, June 2008.

\bibitem{clipping}
J.~Armstrong, ``Peak-to-average power reduction for {OFDM} by repeated clipping
  and frequency domain filtering,'' \emph{Electronics Letters}, vol.~38, no.~5,
  pp. 246--247, Feb 2002.

\bibitem{pts}
S.~H. Muller and J.~B. Huber, ``{OFDM} with reduced peak-to-average power ratio
  by optimum combination of partial transmit sequences,'' \emph{Electronics
  Letters}, vol.~33, no.~5, pp. 368--369, Feb 1997.

\bibitem{SLM}
R.~W. Bauml, R.~F.~H. Fischer, and J.~B. Huber, ``Reducing the peak-to-average
  power ratio of multicarrier modulation by selected mapping,''
  \emph{Electronics Letters}, vol.~32, no.~22, pp. 2056--2057, Oct 1996.

\bibitem{7024183}
J.~Hou, J.~Ge, and F.~Gong, ``Tone reservation technique based on
  peak-windowing residual noise for {PAPR} reduction in {OFDM} systems,''
  \emph{IEEE Trans. Veh. Technol.}, vol.~64, no.~11, pp. 5373--5378, Nov 2015.

\bibitem{taxonomy}
Y.~Rahmatallah and S.~Mohan, ``Peak-to-average power ratio reduction in {OFDM}
  systems: A survey and taxonomy,'' \emph{IEEE Commun. Surveys Tuts.}, vol.~15,
  no.~4, pp. 1567--1592, Mar. 2013.

\bibitem{ICF1}
H.~Chen and A.~M. Haimovich, ``Iterative estimation and cancellation of
  clipping noise for {OFDM} signals,'' \emph{IEEE Commun. Lett.}, vol.~7,
  no.~7, pp. 305--307, July 2003.

\bibitem{ICF2}
Y.~C. Wang and Z.~Q. Luo, ``Optimized iterative clipping and filtering for
  {PAPR} reduction of {OFDM} signals,'' \emph{IEEE Trans. Commun.}, vol.~59,
  no.~1, pp. 33--37, January 2011.

\bibitem{ICF3}
X.~Zhu, W.~Pan, H.~Li, and Y.~Tang, ``Simplified approach to optimized
  iterative clipping and filtering for {PAPR} reduction of {OFDM} signals,''
  \emph{IEEE Trans. Commun.}, vol.~61, no.~5, pp. 1891--1901, May 2013.

\bibitem{Fehri}
B.~Fehri, S.~Boumaiza, and E.~Sich, ``Crest factor reduction of inter-band
  multi-standard carrier aggregated signals,'' \emph{IEEE Trans. Microw. Theory
  Tech.}, vol.~62, no.~12, pp. 3286--3297, Dec 2014.

\bibitem{Traverso}
S.~Traverso, ``A new family of filters for {PAPR} reduction of carrier
  aggregated signals,'' in \emph{Proc. IEEE WCNC}, April 2016, pp. 1--6.

\bibitem{TR}
L.~Wang and C.~Tellambura, ``Analysis of clipping noise and tone-reservation
  algorithms for peak reduction in {OFDM} systems,'' \emph{IEEE Trans. Veh.
  Technol.}, vol.~57, no.~3, pp. 1675--1694, May 2008.

\bibitem{3GPPTS22261}
{3GPP TS 22.261 V16.4.0}, ``{Service requirements for the 5G system; Stage
  1},'' \emph{{\normalfont Tech. Spec. Group Services and System Aspects, Rel.
  16}}, June 2018.

\bibitem{975762}
H.~Ochiai and H.~Imai, ``Performance analysis of deliberately clipped {OFDM}
  signals,'' \emph{IEEE Trans. Commun.}, vol.~50, pp. 89--101, Jan 2002.

\bibitem{J:1987_Sorensen_FFT_complexity}
H.~Sorensen, M.~Heideman, and C.~Burrus, ``{On computing the split-radix
  FFT},'' \emph{IEEE Trans. on Acoust., Speech, and Signal Process.}, vol.~34,
  no.~1, pp. 152--156, February 1986.

\bibitem{NR}
{3GPP TS 38.104 V15.4.0}, ``{New Radio (NR); Base Station (BS) radio
  transmission and reception},'' \emph{{\normalfont Tech. Spec. Group Radio
  Access Network, Rel. 15}}, Dec. 2018.

\end{thebibliography}
%\bibliography{IEEEref}

\end{document}